\documentclass[superscriptaddress, preprint]{revtex4-1}

\usepackage{hyperref}

\usepackage[utf8]{inputenc}
\usepackage{lmodern}
\usepackage[version=3]{mhchem}
\usepackage{amsmath,amssymb,amstext,amsfonts}
\usepackage{natbib}
\usepackage{psfrag,graphicx}
\usepackage[]{units}
\usepackage{upgreek}
\usepackage{textcomp} 
\usepackage{color}
\usepackage{float}
\usepackage{subfigure}
\usepackage{array} 
\usepackage{physics}

\usepackage{epstopdf}

\usepackage[ngerman, num]{isodate}

\begin{document}
\title{Additive Manufactured and Topology Optimized Passive Shimming Elements for Permanent Magnetic Systems}

\author{C. Huber}
\email[Correspondence to: ]{huber-c@univie.ac.at}
\affiliation{Physics of Functional Materials, University of Vienna, 1090 Vienna, Austria}
\affiliation{Christian Doppler Laboratory for Advanced Magnetic Sensing and Materials, 1090 Vienna, Austria}

\author{M. Goertler}
\affiliation{Institute for Surface Technologies and Photonics, Joanneum Research Forschungsgesellschaft GmbH, 8712 Niklasdorf, Austria}

\author{F. Bruckner}
\affiliation{Physics of Functional Materials, University of Vienna, 1090 Vienna, Austria}
\affiliation{Christian Doppler Laboratory for Advanced Magnetic Sensing and Materials, 1090 Vienna, Austria}

\author{C. Abert}
\affiliation{Physics of Functional Materials, University of Vienna, 1090 Vienna, Austria}
\affiliation{Christian Doppler Laboratory for Advanced Magnetic Sensing and Materials, 1090 Vienna, Austria}

\author{I. Teliban}
\affiliation{Magnetfabrik Bonn GmbH, 53119 Bonn, Germany}

\author{M. Groenefeld}
\affiliation{Magnetfabrik Bonn GmbH, 53119 Bonn, Germany}

\author{D. Suess}
\affiliation{Physics of Functional Materials, University of Vienna, 1090 Vienna, Austria}
\affiliation{Christian Doppler Laboratory for Advanced Magnetic Sensing and Materials, 1090 Vienna, Austria}

\date{\today}

$ $\newline

\begin{abstract}
A method to create a highly homogeneous magnetic field by applying topology optimized, additively manufactured shimming elements is investigated. The topology optimization algorithm can calculate a suitable permanent and nonlinear soft magnetic design that fulfills the desired field properties. The permanent magnetic particles are bonded in a polyamide matrix, and they are manufactured with a low-cost, end-user 3D printer. Stray field measurements and an inverse stray field simulation framework can determine printing and magnetization errors. The customized shimming elements are manufactured by a selective melting process which produces completely dense soft magnetic metal parts. The methodology is demonstrated on an example of two axial symmetric cylindrical magnets. In this case, the homogeneity can be increased by a factor of 35. Simulation and measurement results point out a good conformity. 
\end{abstract}

\maketitle

\section*{Introduction}
Maintaining a highly homogeneous magnetic field is a key feature of many magnetic analysis methods and experiments in different scientific fields. Nuclear magnetic resonance (NMR) and magnetic resonance imaging (MRI) are given as examples. A uniform magnetic field is also required in some other applications such as magnetometers, neutron interferometers, magnetic traps, particle counters etc. The resolution of magnetic analysis methods can be improved by generating a stronger and more homogeneous field over the region of interest (ROI). As a result of production tolerances and of the magnetic field of the environment, the magnetic field of a permanent or electromagnetic system will be far from homogeneous compared with an ideal field of the system. The technique to correct the field inhomogeneity is typically called shimming of the magnetic system. In general, two shimming methods exists to increase the magnetic field homogeneity of a permanent magnet. (i) Passive shimming corrects the magnetic field by ferromagnetic materials placed on specific locations along the magnet \cite{ren2009study}. (ii) Active shimming uses electro magnets with specialized coils to generate a correction field \cite{jezzard2006shim, anderson1961electrical}. A passive shimming technique for any kind of permanent magnetic systems is researched in this paper.  

Several optimized permanent magnetic designs exist that obtain a homogeneous magnetic field \cite{marble2007compact, manz2008simple, windt2011portable, raich2004design}. Such magnetic designs can be found with different methods. Examples of numerical optimization methods include: (i) inverse magnetic field computation based on a finite elements method (FEM) where the magnetization $\boldsymbol{M}$ of a defined structure is optimized \cite{inverse_flo, pub_17_1}, (ii) shape optimization improves existing designs for better performance \cite{shape_opt}, (iii) parameter variation simulations can be used to find an optimal layout of predefined magnetic structures \cite{linear, opt_design}, and (iv) topology optimization which allows the designer of magnetic systems to find a suitable topology of the magnets from scratch \cite{topo_motor, pub_17_2, wang2017topology}.

The advantage of topology optimization is the ability to create complicated magnetic field shapes, but this freedom-of-design is also the biggest disadvantage in terms of manufacturing of such optimized structures. Complex, time and cost intensive production processes are necessary. This disadvantage can be eradicated by additive manufacturing (AM) techniques. AM or colloquially called 3D printing is an affordable technique to manufacture models, prototypes, or end-user products with a minimum amount of assigned  material and time.  Recently, it has been shown that an end-user fused deposition modeling (FDM) 3D printer can be used to print polymer-bonded magnets with a complex shape \cite{pub_16_1_apl, pub_17_1, pub_17_2}. The FDM technology works by heating up wire-shaped thermoplastic filaments above the softening point. A movable extruder presses the molten thermoplastic through a nozzle and builds up the object layer by layer \cite{3d-print}. 

Soft magnetic materials can be in-situ synthesized by selective laser melting (SLM). SLM is a powder bed method, implying that objects are created layerwise from metal powder under influence of a localized heat source. After each finished layer, the workpiece is lowered by one layer thickness. Then, a new layer of powder is spread on the top of the object and defined areas are melted selectively by scanning the part's cross-section with a laser beam. Typically, permalloys like FeNi$_3$ or Ni-Fe-V and Ni-Fe-Mo are used, respectively \cite{zhang2013magnetic} \cite{mikler2017laser}. A big advantage of this method is to manufacture dense soft magnetic objects with an arbitrary shape. While the saturation magnetization of these alloys is comparable to conventionally processed versions of similar composition, the coercivities were higher for the laser-processed alloys, presumably due to microstructural defects. Thus, the magnetic properties can be modified by the laser process parameters, that can be used to produce tailored soft magnets for various applications like  transformers, electric motors, and other electromagnetic devices.

In contrast to SLM produced dense magnets, magnets can also be manufactured by FDM from compound materials which consist of soft magnetic particles embedded in a thermoplastic filament. \cite{bollig2017effects}. Samples of commercially available extruded composite filament from Proto-Pasta (Magnetic Iron PLA) are printed. This filament consists of $40$~wt.\% Fe particles embedded in a polylactic acid (PLA) polymer matrix. The magnetic properties of soft magnetic compounds are mainly influenced by their filler fraction. Filling fraction of more than $65$~vol.\% would be necessary to fabricate functional soft magnetic parts. 

This work describes the complete designing and manufacturing method for a magnetic system that generates a homogeneous magnetic field in a defined ROI. With topology and inverse stray field simulation tools, the optimized design can be found. The permanent magnetic structures are printed with a FDM technique. Printing and magnetization errors are considered. A SLM process produces the passive shimming elements for the error correction.

\section*{Method}
\begin{figure}[htbp]
	\centering
	\includegraphics[width=0.5\linewidth]{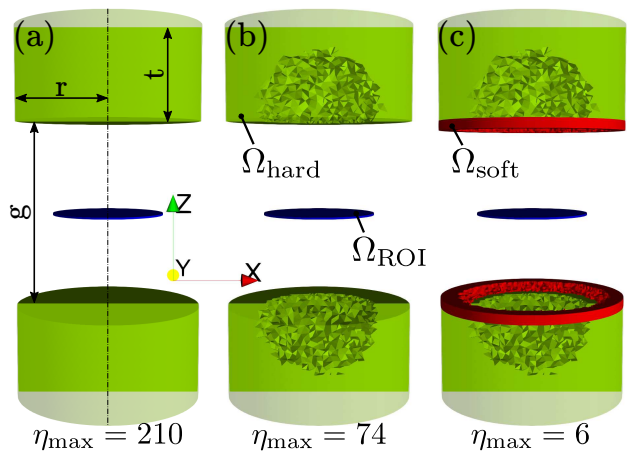}
	\caption{Optimization steps to generate a homogeneous magnetic field in the region $\Omega_\text{ROI}$. (a) Permanent magnetic cylinder magnets ($r=10$~mm, $t=10$~mm) with a gap between the magnets of $g=20$~mm. Magnetization along the $z$-axis. (b) Topology optimized permanent magnetic structure. (c) Printing errors and incorrect magnetization are corrected by topology optimized soft magnetic shimming elements (thickness: $0.5$~mm).}
	\label{fig:iterations}
\end{figure}

The method to find a permanent magnetic design with passive shimming elements that generates a homogeneous magnetic field in a defined ROI should be described by a simple example. The inhomogeneity $\eta$ inside the ROI is defined with the magnetic flux density $B$ as
\begin{equation}
\begin{aligned}
\label{eq:inhomo}
     \eta = \frac{B_\text{max} - B_\text{min}}{B_\text{mean}} .
\end{aligned}
\end{equation}
The ROI has a radius of $6$~mm and a thickness of $1$~mm. The inhomogeneity $\eta$ should be smaller than $10$~\textperthousand . 
As a starting point for the optimization, a naive design of two cylindrical permanent magnets with radius $r=10$~mm, thickness $t=10$~mm, and a gap of $g=20$~mm is chosen. Starting from such a design (Fig.~\ref{fig:iterations}(a)), the topology of this permanent magnetic system should be optimized (Fig.~\ref{fig:iterations}(b)) in oder to minimize the inhomogeneity $\eta$. After magnetization of the system along the $z$-axis, the field is measured and the quality of the print and magnetization is determined by an inverse stray field simulation \cite{pub_17_1, inverse_flo}. The result of this simulation is an input for the topology optimization of the soft magnetic shimming elements (Fig.~\ref{fig:iterations}(c)).

\section*{Simulation Framework}
In a simply connected domain without current, the stray field strength $\boldsymbol{H}$ of a magnetic body is given by
\begin{align}
\boldsymbol{H} = - \nabla u
\end{align}
with the magnetic scalar potential $u$. The relation between the magnetic flux density $\boldsymbol{B}$ and the field strength $\boldsymbol{H}$ is
\begin{align}
\boldsymbol{B} = \mu(\boldsymbol{J}) \boldsymbol{H} + \boldsymbol{J}
\end{align}
with $\mu=\mu_0 \mu_r$. $\mu_0$ is the vacuum permeability and $\boldsymbol{J}=\mu_0 \boldsymbol{M}$ are the magnetic polarization with the magnetization $ \boldsymbol{M}$. The nonlinear isotropic permeability $\mu_r(\boldsymbol{J})$ of the material is defined as
\begin{align}
\mu_r(\boldsymbol{J}) = \frac{\boldsymbol{B(J)}}{\boldsymbol{H}}.
\end{align}

For the topology optimization framework we use a density based method, also known as solid isotropic microstructure with penalization (SIMP) \cite{pub_17_2}. This method is based on a 3D finite-element method (FEM) simulation algorithm. Each tetrahedral finite element has a density parameter $\varrho$, which ranges from $0$ (void) to $1$ (bulk). This leads to only one design parameter per element \cite{topo_trends}. Permanent magnetic systems with a magnetization $\boldsymbol{M}$ of an element in the design domain $\Omega_\text{hard} \in \mathbb{R}^3$ can be formulated for the design method as
\begin{align}
\label{eq:mu}
\boldsymbol{M}(\varrho)=\varrho^k\boldsymbol{M}_0
\end{align}
where $\varrho \in [0,1]$ is the density value of a FEM element, $\boldsymbol{M_0}$ is the constant magnetization vector, and $k=1$ is the penalization parameter \cite{topo_sensor}. For nonlinear isotropic soft magnetic materials, the relative permeability $\mu_r(\boldsymbol{M})$ in the design domain $\Omega_\text{soft} \in \mathbb{R}^3$ can be reformulated for the design method to 
\begin{align}
 \mu_r \left(\boldsymbol{M}(\varrho) \right) = (\mu_{r0}(\boldsymbol{M}) - 1)\varrho^k + 1
\end{align}
for topology optimization of nonlinear isotropic soft magnetic materials with the measured permeability $\mu_{r0}$ , a penalization parameter of $k=4$ leads to good results. 

The general topology optimization problem with the density method can be formulated as
\begin{equation}
\begin{aligned}
\label{eq:min_topo}
   &\text{Find: } \min_{\varrho} J(\varrho) \\
   &\text{subject to: } \int_{\Omega_i} \varrho(\boldsymbol{r})\text{d}\boldsymbol{r}\leq V; \\
   &0 \leq \varrho(\boldsymbol{r}) \leq 1,\,\, \boldsymbol{r} \in \Omega_i
\end{aligned}
\end{equation}
with the objective function $J$ and the maximum Volume $V$ of the design as a constraint; $i\in\{\text{hard},\text{soft}\}$ defines the permanent and soft magnetic domain, respectively.

In our case, to minimize the inhomogeneity of the magnetic flux density $\boldsymbol{B}$ in the ROI, following objective function $J$ has to be minimized
\begin{align}
 J = \int_{\Omega_\text{ROI}} | \grad \otimes \boldsymbol{H} |^2 \mathrm{d} \boldsymbol{r}.
\end{align}

The finite-element package FEniCS is used to implement and solve the demagnetization field problem and the topology optimization method. FEniCS is an open-source software project with the goal to enable automated solution of nonlinear differential equations \cite{AlnaesBlechta2015a}. This involves the automation of: (i) discretization, (ii) discrete solution, (iii) error control, (iv) modeling, and (v) optimization \cite{fenics}. The topology optimization problem can be solved by the adjoint variable method (AVM) \cite{topo_trends}. It is a well-known method for sensitivity analysis using FEM.  The main advantage of this method is the low computational and storage costs compared to other techniques. To solve the topology optimization with the AVM method, the well-suited Dolfin-Adjoint library is used \cite{dolfin, farrell2013automated}. Dolfin-adjoint contains a framework to solve nonlinear partial differential equation (PDE) constraint optimization problems.


\section*{Results}
The topology optimized hard magnetic system should be realized by an FDM 3D printing process \cite{pub_16_1_apl}. A prefabricated compound material (Neofer\textregistered~25/60p) from Magnetfabrik Bonn GmbH is used to realize the setup. It consists of NdFeB particles in a PA11 polymer matrix. The powder has a spherical form, and the NdFeB grains have a uniaxial magnetocrystalline anisotropy, and the orientation of the grains is random leading to isotropic magnetic properties of the bulk magnet. The powder is produced by employing an atomization process followed head treatment. The compound consist of $52$~vol.\% of the magnetic powder.

\begin{figure}[htbp]
	\centering
	\includegraphics[width=0.5\linewidth]{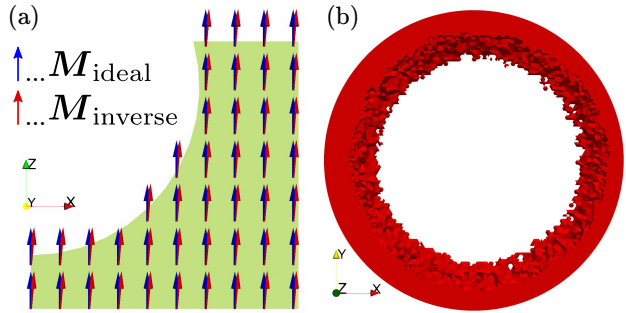}
	\caption{Error correction with passive shimming elements. (a) Cut of the topology optimized magnet of Fig.~\ref{fig:iterations}, where the green area shows the hard magnetic region. Arrows represent the magnetization $\boldsymbol{M}$ of the inverse stray field simulation. The measured magnetic field in the ROI indicates a non-optimal magnetization of the polymer-bonded permanent magnet. (b) Topology optimized soft magnetic shimming elements that correct the magnetization and printing errors.}
	\label{fig:error}
\end{figure}

After printing of the optimized design (Fig.~\ref{fig:iterations}(b)), the objects are magnetized inside an electro magnet with a maximum flux density of $1.9$~T along the $z$-axis. To deduct the quality of the print and the correct magnetization, the field in the ROI is scanned by a 3D stray field measurement setup \cite{pub_16_1_apl}. By the help of an inverse stray field simulation framework, the magnetization distribution inside the permanent magnetic structure is calculated \cite{pub_17_1, inverse_flo}. As shown in Fig.~\ref{fig:error}(a), the magnetization is not perfectly orientated along the $z$-axis. This error originates from a nonconforming magnetization, as well as other printing errors, and it should be eradicated with soft magnetic shimming elements (Fig.~\ref{fig:iterations}(c)). For this reason, the recalculated magnetization is an input for the nonlinear soft magnetic topology optimization. 

The first idea was to use a commercially available soft magnetic compound material (Iron Metal PLA Composite, Proto-Pasta). This compound material shows good printing capability. However, due to its low amount of soft magnetic particles of only $16$~vol.\%, it shows weak magnetic properties compared to complete dense soft magnetic materials.

For bonded soft magnets, the relation between coercivity and filler fraction $\phi_f$ can be described be the equation from Néel \cite{neel1947magnetisme, anhalt2009theoretical}
\begin{align}
 H_c(\phi_f) = H_c(0)(1-\phi_f)
\end{align}
with the coercivity $H_c(0)$ of one isolated magnetic particle of the filler material and the coercivity $H_c(\phi_f)$ for a filler fraction $\phi_f$. This relation leads for $\phi_f=1$ to a coercivity of $H_c(1)=0$~A/m. $H_c$ for soft magnets is small but not zero. Nevertheless, the relation describes the coercivity of bonded soft magnets very well \cite{coercive_soft}.

\begin{figure}[htbp]
	\centering
	\includegraphics[width=0.5\linewidth]{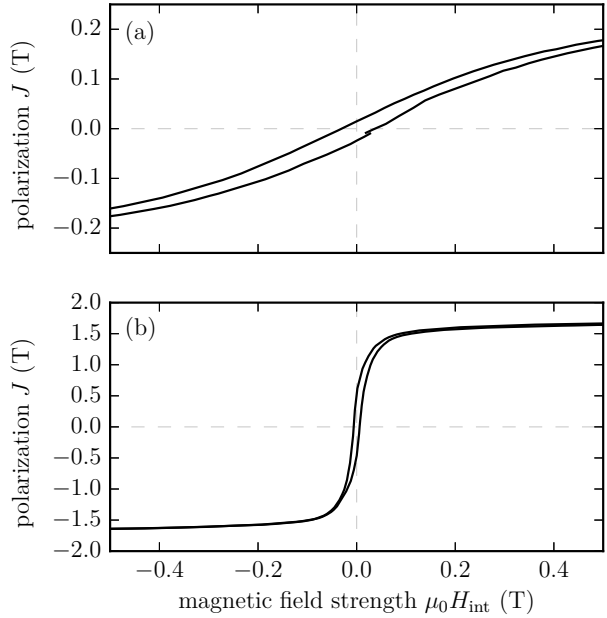}
	\caption{Hysteresis measurements. The measurements are done for cubes of $a=5$~mm and the loops are de-sheared with $N=1/3$.  (a) Polymer-bonded Iron Metal PLA Composite for FDM. (b) EOS MaraginSteel MS1, 1.2709 for SLM.}
	\label{fig:hysteresis}
\end{figure}

The relative permeability $\mu_c$ of soft magnetic compounds can be described by the equation from Bruggeman \cite{bruggeman1935berechnung, soft_compound}
\begin{align}
 \mu_c(\phi_f) = \frac{\mu_m}{(1-\phi_f)^3}
\end{align}
this model assumes that the permeability of the filler material $\mu\rightarrow\infty$ as a basis, as well as that the particles are far away from each other and intersection can be neglected. Therefore, this model is only applicable for low filler fractions ($\phi_f<0.85$). The permeability of the polymer matrix material is $\mu_m\approx 1$. In the case of the Iron Metal PLA Composite from Proto-Pasta, a theoretical permeability of only $\mu=2$ is reachable. This value of the permeability fits well with hysteresis measurement performed by a pulse field magnetometry (PFM) (Hirst PFM11) where the material is printed in a cube shape with edge length of $5$~mm \cite{pfm2, pfm}. All measurements are carried out 3 times with the same parameters - temperature of $297$~K and a magnetic field up to $4$~T peak field. The internal field is $H_{\text{int}}=H_{\text{ext}}-JN/\mu_0$, where $H_{\text{ext}}$ is the external field, $N$ is the average demagnetisation factor for a cube ($N=1/3$) \cite{demag}, and $J$ is the material polarization. Fig.~\ref{fig:hysteresis}(a) shows hysteresis measurements of the Iron Metal PLA Composite for FDM.

Since the permeability of the magnetic material produced by FDM is only around $2$, we decided to utilize a different approach were we expect higher permeability and better properties for the shimming application. As mentioned before it is possible to manufacture soft magnetic materials additively by a SLM process. In our case, the SLM machine EOSINT M280 with a laser power of $400$~W is used. A commercial available steel powder (EOS MaraginSteel MS1, 1.2709) is used for this work \cite{sedlak2015study}. This powder has optimal properties for the SLM process. The printing parameters are summarized in Tab.~\ref{tab:slm_param}.
\begin{table}[htbp]
\centering
\caption{Printing parameters for the SLM of EOS MaraginSteel MS1, 1.2709.}
\label{tab:slm_param}
\begin{tabular}{ll}
paramters       &                    \\ \hline
layer thickness & 40~\textmu m  \\
laser power     & 285~W         \\
scan velocity   & 960~mm/s       \\
hatch distance  & 0.11~mm
\end{tabular}
\end{table}
Fig.~\ref{fig:hysteresis}(b) shows hysteresis measurements of the EOS MaraginSteel MS1 material by SLM. 

The hysteresis curve of the first quadrant is an input parameter for the topology optimization with a nonlinear soft magnetic material (Eq.~\ref{eq:mu}). In combination with the recalculated magnetization distribution, the topology of the soft magnetic shimming elements can be simulated. The maximum radius and thickness of the elements is $6$~mm and $0.5$~mm, respectively. Fig.~\ref{fig:error}(b) shows the simulated topology of the elements. Due to the incorrect magnetization of the permanent magnets, the soft magnetic shimming elements are not rotational symmetric. For a better printing result, the ring shape is divided into two sections. This will avoid poor printing resulting from large overhangs.

The produced shimming elements are then mounted onto the permanent magnets. Fig.~\ref{fig:pic_setup} shows a picture of the fully assembled setup (with shimming elements) during the field measurement. 
\begin{figure}[htbp]
	\centering
	\includegraphics[width=0.5\linewidth]{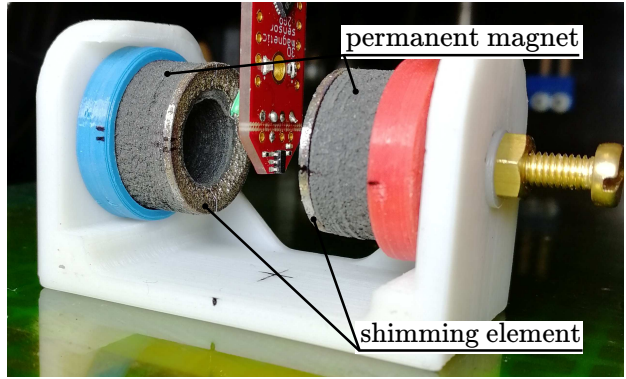}
	\caption{Picture of the setup during the field measurement.}
	\label{fig:pic_setupline_scan}
\end{figure}

\begin{figure}[htbp]
	\centering
	\includegraphics[width=0.5\linewidth]{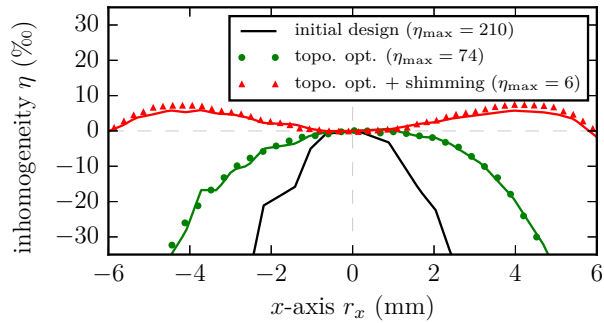}
	\caption{Measurement and simulation of the inhomogeneity $\eta$ of the magnetic flux density along $x$-axis in the middle of the ROI. (Solid lines are simulation results.) }
	\label{fig:line_scan}
\end{figure}

The measurements of the inhomogeneity $\eta$ along the $x$-axis for the different phases of magnet design are shown in Fig.~\ref{fig:line_scan} and are compared to simulation results. Good conformity between simulation and measurement results is given at all different iteration steps. In case of the final design with mounted shimming elements, the maximum inhomogeneity with shimming elements is around $6$~\textperthousand , and therefore it fulfills the design criteria.

\section*{Conclusion}
Additive manufacturing offers new opportunities in the field of magnetic field design and manipulations. It can manufacture objects  with highest individual design flexibility at minimum costs. The full potential of additive manufacturing comes into play when complex and customized parts have to be produced, which would otherwise be complicated to fabricate with conventional subtractive manufacturing methods.

Topology optimization of permanent and nonlinear soft magnetic materials offers the possibility to find a suitable design for a desired field distribution. The disadvantage is that the simulation results are difficult to manufacture. This disadvantage can be eradicate by using additive manufacturing methods. 

Nowadays, 3D printing of polymer-bonded permanent magnetic materials is an active research topic. 3D printing of soft magnetic materials by the aim of a FDM technique is not particular due to its low filler fraction of soft magnetic powder. For this reason, a SLM process is a better technique to manufacture soft magnets for field shaping applications. Nevertheless, if only a weak modification of the external field is necessary (field inhomogeneity in the range of parts per million), FDM with bonded soft magnetic materials could be suitable to shape the field in a small range. 

A homogeneous magnetic field is necessary for many experiments and magnetic analysis methods. Traditionally, shimming elements of simple geometric shape are used to minimize field inhomogeneities. This work presents a method to find a proper topology optimized design that generate a homogeneous magnetic field in a defined region. The inhomogeneity can be decrease by a factor of around $35$. Even more, unavoidable printing and magnetization errors can be detected by an inverse stray field simulation technique which shows good accordance to the measured data. These errors can be considered in the next iteration step.

\section*{Acknowledgments}
The support from CD-Laboratory AMSEN (financed by the Austrian Federal Ministry of Economy, Family and Youth, the National Foundation for Research, Technology and Development) is acknowledged. The authors would like to thanks Montanuniversitaet Leoben for the extrusion of the filaments. The computational results presented have been achieved using the Vienna Scientific Cluster (VSC).

\bibliographystyle{aipnum4-1}

%

\end{document}